\documentclass[aps,prl,letterpaper,reprint,preprintnumbers]{revtex4-1}

\usepackage{amsmath,amsfonts,amssymb}
\usepackage{bbm}
\usepackage{xcolor}
\usepackage{graphicx}
\usepackage[bookmarks,bookmarksnumbered]{hyperref}

\newcommand{\bmat}{\left(\begin{array}}
\newcommand{\emat}{\end{array}\right)}
\newcommand{\be}{\begin{equation}}
\newcommand{\ee}{\end{equation}}
\newcommand{\ba}{\begin{eqnarray}}
\newcommand{\ea}{\end{eqnarray}}

\def\lsim{\raise0.3ex\hbox{$\;<$\kern-0.75em\raise-1.1ex\hbox{$\sim\;$}}}
\def\gsim{\raise0.3ex\hbox{$\;>$\kern-0.75em\raise-1.1ex\hbox{$\sim\;$}}}

\def\be{\beta}

\begin{document}

\preprint{HIP-2015-39/TH}

\title{Higgs--inflaton coupling from reheating and the metastable Universe}

\author{Christian Gross}
\author{Oleg Lebedev}
\author{Marco Zatta}
\affiliation{Department of Physics and Helsinki Institute of Physics, Gustaf H\"allstr\"omin katu 2, FI-00014 Helsinki, Finland}

\begin{abstract}
Current Higgs boson and top quark data favor metastability of our vacuum which raises questions as to why the Universe has chosen an energetically disfavored state and remained there during inflation. In this Letter, we point out that these problems can be solved by a Higgs--inflaton coupling which appears in realistic models of inflation. Since an inflaton must couple to the Standard Model particles either directly or indirectly, such a coupling is generated radiatively, even if absent at tree level. As a result, the dynamics of the Higgs field can change dramatically.
\end{abstract}

%\pacs{pacs numbers}

\maketitle

%%%%%%%%%%%%%%%%%%%%%%%%%%%%%%%%%%%%%%%%%%%%
The current Higgs mass $m_h= 125.15 \pm 0.24 $ GeV and the top quark mass $m_t = 173.34 \pm 0.76 \pm 0.3$ GeV indicate that in the Standard Model (SM) the Higgs quartic coupling turns negative at high energies implying metastability of the electroweak (EW) vacuum at 99\% CL~\cite{Buttazzo:2013uya}. The (much deeper) true minimum of the scalar
potential appears to be at very large field values.
In the cosmological context, this poses a pressing question why the Universe has chosen an energetically disfavored state and why it remained there during inflation despite quantum fluctuations.

In this Letter, we argue that these puzzles can be resolved by a Higgs--inflaton coupling~\cite{Lebedev:2012sy} which appears in realistic models of inflation. Indeed, the energy transfer from the inflaton to the SM fields necessitates interaction between the two in some form. This in turn induces a Higgs--inflaton coupling via quantum effects, even if it is absent at tree level. We find that the loop induced coupling can be sufficiently large to make a crucial impact on the Higgs field evolution.

Another factor that can affect the Higgs field dynamics is the non--minimal scalar coupling to gravity, which creates an effective mass term for the Higgs field~\cite{Espinosa:2007qp,Herranen:2014cua}.
 Here we assume such a coupling to be negligible. 
The effect of quantum fluctuations during inflation has recently been considered in~\cite{Espinosa:2015qea,Hook:2014uia}. 
The conclusion is that the Hubble rate $H$ above the Higgs instability scale leads to destabilization of the EW vacuum, which poses a problem for this class of inflationary models. Related issues have been studied in~\cite{Fairbairn:2014zia,Shkerin:2015exa,Kamada:2014ufa}. 

The Higgs potential at large field values is approximated by~\cite{Altarelli:1994rb}
\begin{equation}
 V_h \simeq {\lambda_{h} (h) \over 4} ~ h^4 \;,
\end{equation}
where we have assumed the unitary gauge $H^T = (0, h/\sqrt{2})$ and 
$\lambda_{h} (h)$ is a logarithmic function of the Higgs field. The current data 
indicate that $\lambda_{h}$ turns negative at around $10^{10}$ GeV~\cite{Buttazzo:2013uya}, although the uncertainties are still significant. 
In the early Universe, the Higgs potential is modified by the Higgs--inflaton coupling
$V_{h\phi}$ with the full scalar potential being 
\begin{equation}
V= V_h + V_{h\phi}+ V_{\phi} \;,
\end{equation}
where $V_{\phi}$ is the inflaton potential.
Since the inflaton must couple to the SM fields either directly or through mediators as required by
successful reheating, quantum corrections induce a Higgs-inflaton
interaction.

In what follows, we consider a few representative examples of reheating models. 
We focus on the Higgs couplings to the inflaton $\phi$ which are {\it required} by renormalizability of the model. Such couplings are induced radiatively with divergent coefficients
and necessitate the corresponding counterterms.
The dim-4 Higgs--inflaton interaction 
takes the form
\begin{equation}
V_{h\phi}= {\lambda_{h\phi} \over 4} ~ h^2 \phi^2 + {\sigma_{h\phi} \over 2} ~ h^2 \phi \;,
\end{equation}
where $\lambda_{h\phi}$ and $ \sigma_{h\phi}$ are model--dependent couplings. 
As we show below, the range of $\lambda_{h\phi}$ relevant to the Higgs potential stabilization is between $10^{-10}$ and $10^{-6}$ (see also~\cite{Lebedev:2012sy}).
For definiteness, we choose a quadratic inflaton potential~\cite{Linde:1983gd} as a representative example
of large field inflationary models,
\begin{equation}
V_{\phi}= {m^2\over 2} ~ \phi^2 + \Delta V_{\rm 1-loop} \;,
\end{equation}
where $m \simeq 10^{-5} M_{\rm Pl}$ and 
$\Delta V_{\rm 1-loop}$ is the radiative correction generated by various couplings of the model. We require this correction to be sufficiently small such that the predictions for cosmological observables of the $\phi^2$--model are not affected, although some
quantum effects can be beneficial~\cite{Enqvist:2013eua}. 
The divergent contributions to $\Delta V_{\rm 1-loop}$ are renormalized in the usual fashion
and the result is given by the Coleman--Weinberg potential~\cite{Coleman:1973jx}.
The leading term at large $\phi$ is the quartic coupling 
\begin{equation}
 \Delta V_{\rm 1-loop} \simeq {\lambda_{\phi} (\phi) \over 4} ~ \phi^4 \;,
\end{equation} 
 with $\lambda_{\phi}$ being logarithmically dependent on $\phi$. 

The energy transfer from the inflaton to the SM fields in general proceeds both through non--perturbative effects and perturbative inflaton decay~\cite{Kofman:1997yn,Allahverdi:2010xz}.
In what follows, we make the simplifying assumption that the reheating is dominated by the perturbative inflaton decay such that the reheating temperature is given by  $T_R \simeq 0.2 \sqrt{\Gamma M_{Pl}}$, where $\Gamma$ is the inflaton decay rate. While this assumption is essential for establishing a correlation between $\lambda_{h\phi}$ and $T_R$, it does not affect the range of  $\lambda_{h\phi}$ consistent with the inflationary predictions.  
We consider three representative reheating scenarios which assume no tree level interaction between the Higgs and the inflaton, and compute the consequent loop--induced couplings.

\begin{figure}[t] 
\centering{
\includegraphics[scale=0.51]{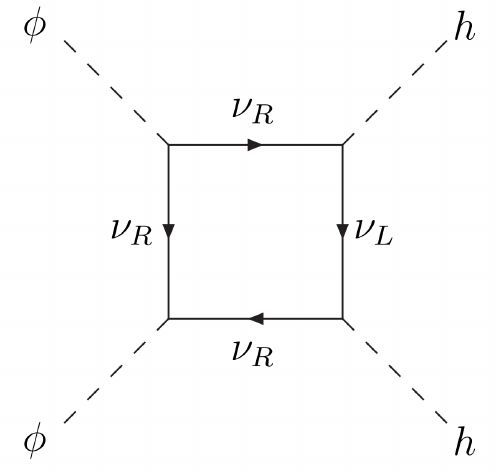}
\includegraphics[scale=0.52]{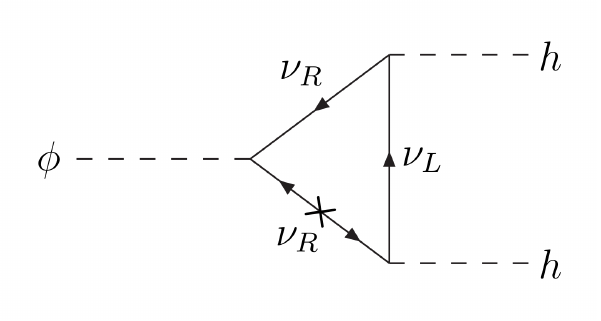}
\includegraphics[scale=0.51]{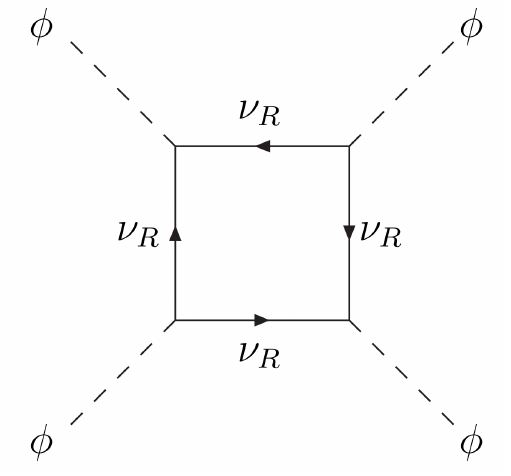}
}
\caption{ \label{plots1}
Leading radiatively induced scalar couplings via the right--handed neutrinos. (Diagrams with the same topology are not shown). 
}
\end{figure}

%%%%%%%%%%%%%%%%%%%%%%%%%%%%%%%%%%%%%%%%%%%%
{\bf 1.~Reheating via right--handed neutrinos.} The inflaton energy is transferred to the SM sector via its decay into right--handed Majorana neutrinos $\nu_R$ which in turn
produce SM matter. The added benefit of this model is that the heavy
neutrinos may also be responsible for the matter--antimatter asymmetry of the Universe via leptogenesis~\cite{Fukugita:1986hr}.
The relevant tree level Lagrangian reads 
\begin{equation}
-\Delta {\cal L} = {\lambda_\nu \over 2} ~\phi \nu_R \nu_R + {y_\nu } ~
\bar l_L \! \cdot \! H^* \, \nu_R + {M\over 2} ~ \nu_R \nu_R + {\rm h.c.}~, 
\end{equation}
where $l_L$ is the lepton doublet, $M$ is chosen to be real and 
we have assumed that a single $\nu_R$ species dominates.
 These interactions generate a coupling between the Higgs and the inflation at 1 loop 
 (Fig.~\ref{plots1}). Since we are interested in the size of the radiatively induced couplings,
let us impose the renormalisation condition that they vanish at a given high energy scale, say the 
Planck scale $M_{\rm Pl}=2.4 \times 10^{18}$ GeV. Then, a finite correction is induced at the scale 
relevant to the inflationary dynamics, which we take to be the Hubble rate $H= m \phi /\left(\sqrt{6} 
M_{\rm Pl}\right)$, with other choices leading to similar results. 
We find in the leading--log approximation,
\begin{eqnarray}
\lambda_{h\phi}&\simeq& { \vert \lambda_\nu y_\nu \vert^2 \over 2 \pi^2} \ln {M_{\rm Pl}\over H} \;, 
\nonumber\\
\sigma_{h\phi}&\simeq& -{ M \vert y_\nu \vert^2 {\rm Re} \lambda_\nu \over 2 \pi^2} \ln {M_{\rm Pl}\over 
H} \;, \nonumber \\
\lambda_{\phi}&\simeq& { \vert \lambda_\nu \vert^4 \over 4 \pi^2} \ln {M_{\rm Pl}\over H} \;.
\end{eqnarray}
Here we have chosen the same renormalization condition for $\lambda_\phi$ and $\lambda_{h\phi}, 
\sigma_{h\phi}$. Since the dependence on the renormalization scale is only logarithmic, this assumption 
does not affect our results.
The most important constraint on the couplings is imposed by the inflationary predictions.
Requiring $ \lambda_{\phi} \phi^4/4 \ll m^2 \phi^2/2$ in the last 60 $e$-folds of expansion
(see e.g.~\cite{Lyth:1998xn}), we find $\lambda_\phi \ll 2 \times 10^{-12}$ and therefore
$ \lambda_\nu < 1 \times 10^{-3}$. The seesaw mechanism also limits the size of the Yukawa coupling 
$y_\nu$.The experimental constraints on the mass of the active neutrinos require
approximately $(y_\nu v)^2/M<1$ eV. 
Assuming that the perturbative decay of the inflaton dominates, the mass of the right-handed neutrinos
is bounded by $M<10^{13}$ GeV, which in turn implies $y_\nu<0.6$.
We therefore get an upper bound on the 
size of the Higgs--inflaton coupling,
\begin{equation}
\lambda_{h\phi} < 2 \times 10^{-7} \;.
\end{equation}
Note that $\lambda_{h\phi} $ is positive and thus the inflaton creates a positive effective mass term for 
the Higgs. The trilinear $\phi h^2$ term is irrelevant as long as 
$\vert \lambda_\nu\vert \phi \gg M$, which is the case for all interesting applications. (Similarly,
the cubic term $\phi^3$ is negligible.)

During the inflaton oscillation stage, the magnitude of $\phi$ decreases as $1/t$. 
When the effective masses of $\nu_R$ and $h$ turn sufficiently small, the decays $\phi
\rightarrow \nu_R \nu_R$, $\phi \rightarrow h h $ become allowed. The constraints above imply
$\Gamma(\phi \rightarrow \nu_R \nu_R ) \gg \Gamma(\phi \rightarrow hh) $ and therefore 
the total inflaton decay width is
$\Gamma = { \vert \lambda_\nu \vert^2 \over 32 \pi } ~m ,$
where we have neglected the $\nu_R$ mass compared to that of the inflaton. Assuming 
that the right--handed neutrinos decay promptly and the products thermalize (or $\nu_R$
themselves thermalize)
 so that $T_R \simeq 0.2 \sqrt{\Gamma M_{\rm Pl}}$,
we find the following correlation between the Higgs--inflaton coupling and the reheating temperature $T_R$,
\begin{equation}
\lambda_{h\phi} 
\simeq 5 \times 10^{-7} \; \vert y_\nu \vert^2 ~\left( {T_R \over 1.5 \times 10^{11}
\;{\rm GeV} } \right)^2 ~,
\end{equation}
where $T_R$ is bounded by $1.5 \times 10^{11}$ GeV. Note that this relation holds only
under the assumption of perturbative reheating. Therefore, for the neutrino Yukawa coupling and the reheating temperature within one--two orders of magnitude from their upper bounds, the dynamics of the Higgs evolution change drastically. 
Similar conclusions apply to models with multiple $\nu_R$ species.

\begin{figure}[t] 
\centering{
\includegraphics[scale=0.51]{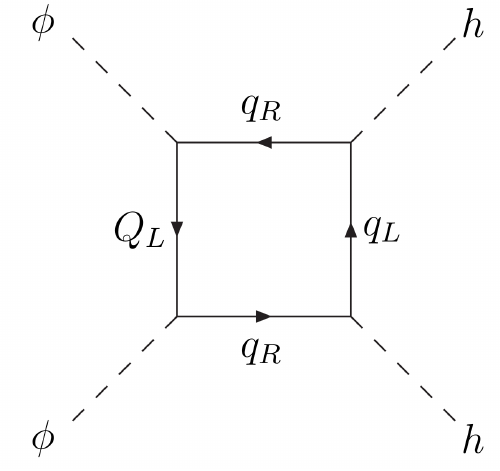}
\includegraphics[scale=0.52]{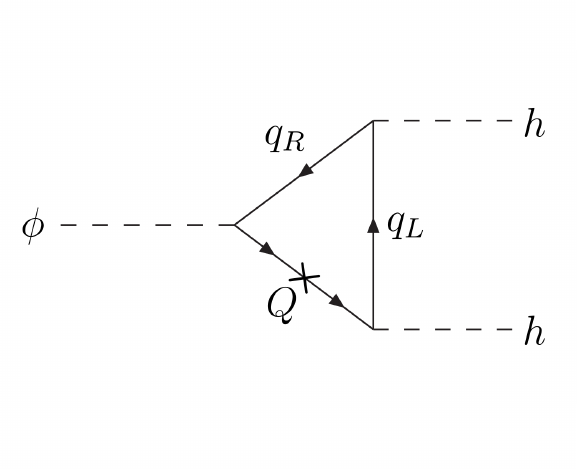}
\includegraphics[scale=0.51]{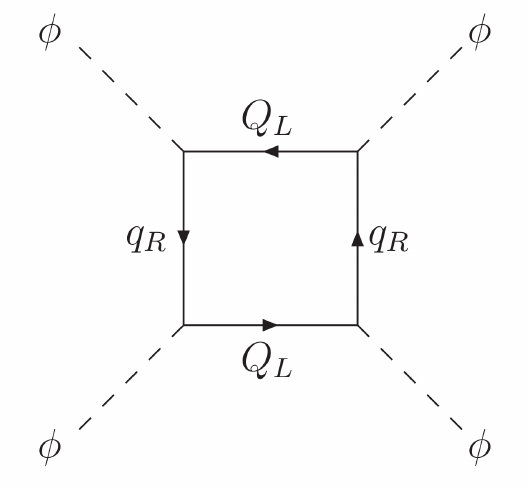}
}
\caption{ \label{plots2}
Leading radiatively induced scalar couplings via the vector--like quarks $Q_{L,R}$
and SM quarks $q_{L,R}$. (Diagrams with the same topology are not shown). 
}
\end{figure}

%%%%%%%%%%%%%%%%%%%%%%%%%%%%%%%%%%%%%%%%%%%%
{\bf 2.~Reheating and non--renormalizable operators.}
A common approach to reheating is to assume the presence 
of non--renormalizable operators that couple the inflaton to the SM fields. Let us consider 
a representative example of the following operators
\begin{equation}
O_1 = {1\over \Lambda_1} \phi\; \bar q_L \! \cdot \!  H^* \, t_R ~~,~~
O_2 = {1\over \Lambda_2} \phi\; G_{\mu\nu} G^{\mu\nu} ~, 
\end{equation} 
where $\Lambda_{1,2}$ are some scales, $G_{\mu\nu}$ is the gluon field strength and $q_L,t_R$ are the third generation quarks. These couplings allow for a direct decay of the inflaton into the SM particles.
It is again clear that a Higgs--inflaton interaction is induced radiatively. In order to calculate the 1--loop couplings reliably, one needs to complete the model in the ultraviolet (UV). The simplest possibility to obtain an effective dim-5 operator is to integrate out a heavy fermion. Therefore, we introduce vector--like quarks $Q_L, Q_R$ with the tree level interactions
\begin{equation}
-\Delta {\cal L} = {y_Q} \; \bar q_L \! \cdot \!  H^* \, Q_R + {\lambda_Q } \; \phi \;
\bar Q_L t_R + {\cal M} \; \bar Q_L Q_R + {\rm h.c.}~, 
\end{equation}
where the heavy quarks have the quantum numbers of the right--handed top $t_R$,
${\cal M}$ is above the inflaton mass and the couplings to the third generation are assumed to dominate.
One then finds that $O_1$ appears at tree level with
$1/\Lambda_1 = y_Q \lambda_Q/{\cal M}$, whereas $O_2$ appears only at 2 loops 
with $1/\Lambda_2 \sim y_Q \lambda_Q y_t \alpha_s/(64 \pi^3 {\cal M})$ and
can be neglected. Using the renormalization condition that the relevant couplings vanish at the Planck scale, we get in the leading--log
approximation (see Fig.~\ref{plots2})
\begin{eqnarray}
\lambda_{h\phi}&\simeq& { 3 \vert \lambda_Q y_t \vert^2 \over 2 \pi^2} \ln {M_{\rm Pl}\over {\cal M}} \;, \nonumber\\
\sigma_{h\phi}&\simeq& -{ 3 {\cal M} \;{\rm Re} (\lambda_Q y_Q y_t) \over 2 \pi^2} \ln {M_{\rm Pl}\over {\cal M}} \;, \nonumber \\
\lambda_{\phi}&\simeq& { 3 \vert \lambda_Q \vert^4 \over 2 \pi^2} \ln {M_{\rm Pl}\over {\cal M}} \;,
\end{eqnarray}
where $y_t$ is the top Yukawa coupling and
we have assumed ${\cal M}\ll M_{\rm Pl}$. Requiring smallness of the correction to the inflaton potential in the last 60 $e$--folds, we 
get $\vert \lambda_Q \vert < 2 \times 10^{-3}/ (\ln \;M_{\rm Pl}/{\cal M} )^{1/4}$
and obtain the bound
\begin{equation}
\lambda_{h\phi} < 10^{-7} \; \left( \ln {M_{\rm Pl}\over {\cal M}}
\right)^{1/2} ~,
\end{equation}
where we have taken $y_t({\cal M}) \simeq 0.5$. 
For ${\cal M}$ in the allowed range, this implies $\lambda_{h\phi} < 3 \times 10^{-7}$. We find again that $\lambda_{h\phi} $ is positive and can be large enough to affect the Higgs evolution. Assuming no large hierarchy between $\lambda_Q$ and $y_Q$,
we have $\phi \vert\lambda_Q\vert \gg {\cal M} \vert y_Q \vert$ and the trilinear $\phi h^2$
term is unimportant for the Higgs evolution.

The trilinear interaction is however important for the inflaton decay.
Taking for simplicity the couplings to be real, we have $\Gamma (\phi \rightarrow tth)= \lambda_Q^2 y_Q^2 m^3/(512 \pi^3 {\cal M}^2)$ and $\Gamma (\phi \rightarrow hh)= \sigma_{h\phi}^2/(32 \pi m)$, which implies
\begin{equation}
{\Gamma (\phi \rightarrow tth) \over \Gamma (\phi \rightarrow hh)}
= {\pi^2 \over 36 y_t^2 \; (\ln \;M_{\rm Pl}/{\cal M} )^2} \; {m^4 \over {\cal M}^4}
\ll 1 
\end{equation}
 even for ${\cal M}$ just above the inflaton mass. Therefore the radiatively induced 
 coupling dominates the inflaton decay. (This conclusion can be avoided by tuning the phases of $\lambda_Q$ and $y_Q$ such that Re$(\lambda_Q y_Q) \simeq 0$.)

 Due to the above constraints, the reheating temperature is bounded by $T_R < 10^{-3} {\cal M} \vert y_Q \vert$ $
\left( \ln \;M_{\rm Pl}/{\cal M}
\right)^{3/4} $ for real couplings. Taking $\vert \lambda_Q \vert M_{\rm Pl} $ as the upper bound on $ \vert y_Q \vert {\cal M} $ (see above) and allowing for the maximal value of ${\cal M}$ to be 
$10^{-2} M_{\rm Pl}$, one finds $T_R < 5 \times 10^{12}$ GeV.
An approximate correlation between $\lambda_{h\phi}$ and $T_R$ 
can be expressed as 
\begin{equation}
\lambda_{h\phi} \simeq 10^{-1} ~{\vert\lambda_Q\vert \over \vert y_Q\vert } ~{T_R \over {\cal M}} \;.
\end{equation}

%%%%%%%%%%%%%%%%%%%%%%%%%%%%%%%%%%%%%%%%%%%%
{\bf 3.~Reheating through dark matter production.} This somewhat more exotic scenario
exhibits different qualitative features. It 
assumes that the inflaton interacts mostly with dark matter or some other SM singlet, which
then produces the SM fields through rescattering. The simplest renormalizable model of this type is based on scalar DM $s$ with the tree level 
interactions
\begin{equation}
-\Delta {\cal L} = {\lambda_{\phi s} \over 4} \phi^2 s^2 +
{\sigma_{\phi s} \over 2 } \phi s^2 + {\lambda_{hs}\over 4} h^2 s^2 + {\lambda_{s}\over 4} s^4 +{m_s^2 \over 2} s^2\;.
\end{equation}
In this case, DM is produced both through the non--perturbative effects
and inflaton decay, while the SM particles are generated via the Higgs field. Assuming that DM is much lighter than the inflaton,
 the induced scalar couplings in the leading--log approximation are
\begin{eqnarray}
\lambda_{h\phi}&\simeq& -{ \lambda_{\phi s} \lambda_{hs} \over 16 \pi^2 } \ln {M_{\rm Pl}\over H} \;, \nonumber\\
\sigma_{h\phi}&\simeq& -{ \lambda_{hs} \sigma_{\phi s} \over 16 \pi^2} \ln {M_{\rm Pl}\over H} \;, \nonumber \\
\lambda_{\phi}&\simeq& -{\lambda_{\phi s}^2 \over 32 \pi^2 } \ln {M_{\rm Pl}\over H} \;.
\end{eqnarray}

\begin{figure}[b] 
\centering{
\includegraphics[scale=0.425]{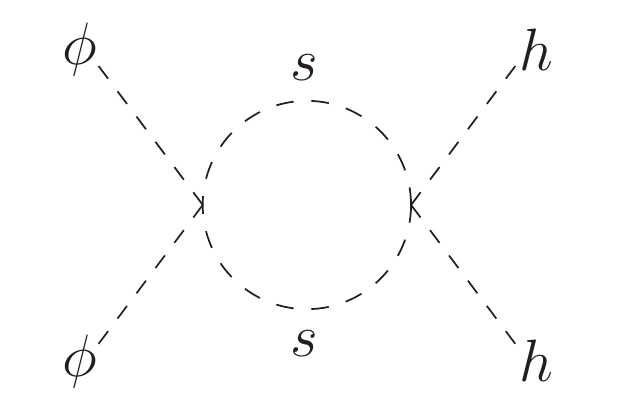}
\includegraphics[scale=0.425]{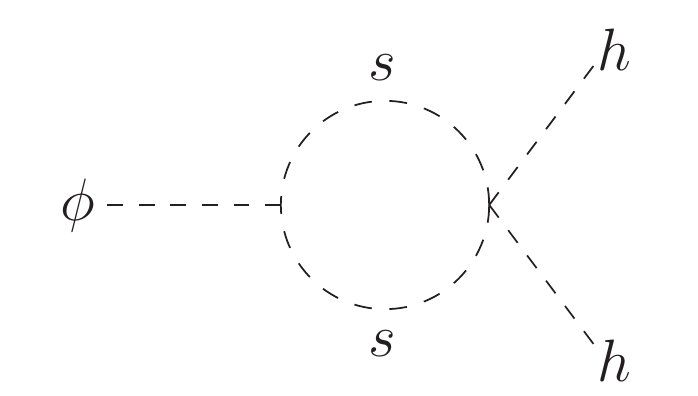}
\includegraphics[scale=0.425]{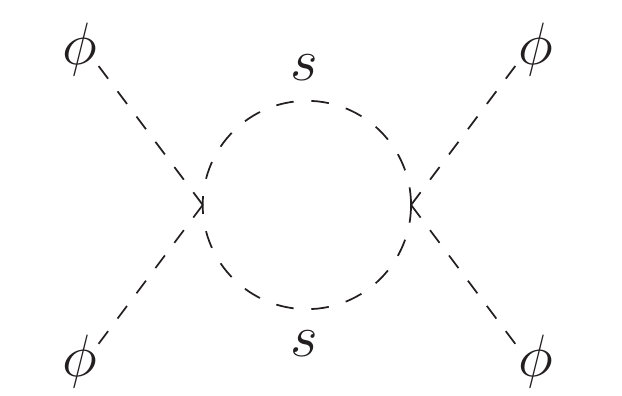}
}
\caption{ \label{plots3}
Leading radiatively induced scalar couplings via scalar dark matter. 
}
\end{figure}

Unlike in the previous examples, we see that $\lambda_{h\phi}$ can be of either sign.
It is positive for $ \lambda_{\phi s} \lambda_{hs} <0$, which is an admissible possibility.
The $\phi^4$ interaction gives a small contribution to the inflaton potential for $\vert \lambda_{\phi s}\vert < 8 \times 10^{-6}$, which implies
\begin{equation}
\vert \lambda_{h\phi} \vert < 5\times 10^{-7} \; \vert \lambda_{hs} \vert ~.
\end{equation}
Here $\lambda_{hs}$ is only restricted by perturbativity and can be as large as ${\cal O
}(1)$ which results in even more significant inflaton--Higgs coupling than before.
The trilinear term is unimportant for the Higgs field evolution for 
$ \lambda_{\phi s} \phi \gg \sigma_{\phi s}$.
Note that since the inflaton decay proceeds mostly through the $\sigma_{\phi s}$ coupling, at leading--log level there is no connection between the reheating temperature and the size
of the induced $\lambda_{h\phi}$. Finally, the model at hand can be viewed as a template
for a class of models which involve a scalar mediator between the inflaton and the SM or dark matter.

 The above examples show that a sizeable $\lambda_{h\phi}$ can generally be induced in 
 realistic reheating models. It can therefore make a crucial impact on the Higgs 
 field evolution. 
 Consider the typical situation that the trilinear $\phi h^2$ term is small compared to the quartic $\phi^2 h^2 $ interaction. With positive $\lambda_{h \phi}$,
the Higgs potential $V_h + V_{h\phi}$ is positive for 
\begin{equation} 
 \phi > \sqrt{\vert \lambda_h \vert \over \lambda_{h \phi} } \; h ~.
 \label{phi-bound}
\end{equation} 
At larger inflaton values, the Higgs potential is convex and dominated by the Higgs--inflaton interaction term which creates an effective Higgs mass $m_h = 
\phi \sqrt{ \lambda_{h \phi}/2}$. If such initial conditions are created and the effective mass is sufficiently large, the Higgs field evolves to zero.

In the reheating models above, we have obtained the upper bound $\lambda_{h \phi} < 10^{-6}$ with some model--dependent variations. Using $\vert \lambda_h \vert \simeq 10^{-2}$
at energies far above the instability scale $10^{10}$ GeV~\cite{Buttazzo:2013uya},
we find that the initial value of the inflaton $\phi_0$ must exceed that of the Higgs
field $h_0$ by at least two orders of magnitude. The use of our renormalizable Higgs potential is meaningful as long as $h_0 \ll M_{\rm Pl}$ so that in practice we take $0.1\ M_{\rm Pl}$ as the upper bound on $h_0$. In that case, the minimal value of $\phi_0 $ is about $10 \ M_{\rm Pl}$, which is typical for large--field inflation models.

The evolution of the system at large field values is governed by the equations
\begin{equation}
 \ddot{h} +3H \dot{h} + {\partial V \over \partial h} =0~~,~~
 \ddot{\phi} +3H \dot{\phi} + {\partial V \over \partial \phi} =0~~,
\end{equation}
where $3 H^2 M_{\rm Pl}^2 = \dot h^2/2 + \dot \phi^2/2 +V$ and $V \simeq m^2 \phi^2/2 + 
\lambda_{h \phi} h^2\phi^2/4$. Taking the initial values of $\dot h$ and $\dot \phi$
to be small, we find the following hierarchy
\begin{equation}
m_\phi \ll H \ll m_h ~,
\end{equation}
where the effective inflaton mass is $m_\phi= \sqrt{m^2 + \lambda_{h \phi} h^2/2} $. 
Therefore, the Higgs field evolves quickly while the inflaton undergoes the usual slow
roll. The magnitude of $h$ decreases linearly,
$ h \sim (\cos m_h t)/m_h t $, and within a few Hubble times $H^{-1}$ the Higgs field value reduces by an order of magnitude~\cite{Lebedev:2012sy}. 
After that the Hubble rate is dominated by the inflaton mass term $H \simeq 
m \phi /\left(\sqrt{6} M_{\rm Pl}\right)$ and the usual slow roll inflation begins.
Since the effective mass of the Higgs field is large and approximately constant,
it evolves exponentially quickly to zero,
\begin{equation}
\vert h(t) \vert \sim e^{-{3\over 2} Ht } \vert h(0) \vert \;.
\end{equation}
After 20 $e$--folds it becomes of electroweak size. This mechanism is operative as long
as $m_h>3H/2$ such that the allowed range of $\lambda_{h \phi}$ is
\begin{equation}
10^{-10} < \lambda_{h \phi} < 10^{-6} \;.
\end{equation}
In this range, the quantum fluctuations of $h$ during inflation are also insignificant 
since (i) the Higgs field is heavy and (ii) the barrier separating the two vacua
is at large field values $h_{\rm bar} \sim \sqrt{ \lambda_{h \phi} \over \vert \lambda_h\vert} \phi \gg H$. The lower bound on $\lambda_{h \phi}$ also guarantees
that the classical evolution of $\phi$ dominates, i.e. 
the initial inflaton value satisfies $\phi/M_{\rm Pl} < 5/\sqrt{m/M_{\rm Pl}}$~\cite{Linde:2005ht}.
The total number of $e$--folds is about $(\phi_0/M_{\rm Pl})^2/4$, with $\phi_0$
bounded by Eq.~(\ref{phi-bound}).

The presence of a small trilinear term $\phi h^2$ does not affect these considerations. 
As long as the effective Higgs mass term remains large and positive, the Higgs field
evolves to zero. In that case, its effect is negligible. 
 The Higgs--inflaton interaction offers no solution to the cosmological 
problems if the effective Higgs mass term is too small or negative.
In that case, $h$ is overwhelmingly likely to end up in the catastrophic true vacuum. 

Since we introduce additional fields that couple to the Higgs,
one may wonder how those affect the running of the Higgs quartic coupling. In the first two examples, this effect is small since the
extra states are very heavy and the (negative) leading contribution to
the beta--function is proportional to the fourth power of the
Higgs--fermion coupling. In the case of scalar mediators, the effect
can be significant depending on the scalar mass and its coupling
to the Higgs. For $m_s \sim$ TeV and $\lambda_{hs}(H) \gsim 0.6$,
the Higgs potential is stable up to the Planck scale (see e.g.~\cite{Lebedev:2012zw}). In that case,
the cosmological problems discussed in this Letter do not arise.
However, for heavier $m_s$ and/or smaller couplings the electroweak vacuum is still metastable, while the stabilization mechanism described here is at work.

In summary, reheating the Universe after inflation necessitates (perhaps indirect) interaction between the inflaton and the SM fields.
As a result, a Higgs--inflaton coupling is induced radiatively as required by renormalizability of the model. Such a coupling can be sufficiently large to alter drastically the Higgs field dynamics in the early Universe. In particular, it can hold the key to the question how the Universe has evolved to the energetically disfavored state, given that the current data point to metastability of the electroweak vacuum.

\begin{acknowledgments}
{\bf Acknowledgments} This work was supported by the Academy of Finland, project ``The Higgs boson and the Cosmos''.
\end{acknowledgments}


\begin{thebibliography}{99}

%\cite{Buttazzo:2013uya}
\bibitem{Buttazzo:2013uya} 
 D.~Buttazzo, G.~Degrassi, P.~P.~Giardino, G.~F.~Giudice, F.~Sala, A.~Salvio and A.~Strumia,
 %``Investigating the near-criticality of the Higgs boson,''
 JHEP {\bf 1312}, 089 (2013).
 %%CITATION = ARXIV:1307.3536;%%
 %292 citations counted in INSPIRE as of 14 Jun 2015

%\cite{Lebedev:2012sy}
\bibitem{Lebedev:2012sy} 
 O.~Lebedev and A.~Westphal,
 %``Metastable Electroweak Vacuum: Implications for Inflation,''
 Phys.\ Lett.\ B {\bf 719}, 415 (2013).
 %%CITATION = ARXIV:1210.6987;%%
 %28 citations counted in INSPIRE as of 04 Jun 2015

%\cite{Espinosa:2007qp}
\bibitem{Espinosa:2007qp} 
 J.~R.~Espinosa, G.~F.~Giudice and A.~Riotto,
 %``Cosmological implications of the Higgs mass measurement,''
 JCAP {\bf 0805}, 002 (2008).
 %%CITATION = ARXIV:0710.2484;%%
 %124 citations counted in INSPIRE as of 14 juin 2015

%\cite{Herranen:2014cua}
\bibitem{Herranen:2014cua} 
 M.~Herranen, T.~Markkanen, S.~Nurmi and A.~Rajantie,
 %``Spacetime curvature and the Higgs stability during inflation,''
 Phys.\ Rev.\ Lett.\ {\bf 113}, no. 21, 211102 (2014)
 %%CITATION = ARXIV:1407.3141;%%
 %22 citations counted in IN
 \textit{and} arXiv:1506.04065 [hep-ph].
 %%CITATION = ARXIV:1506.04065;%%
 
 
 %\cite{Espinosa:2015qea}
\bibitem{Espinosa:2015qea}
  J.~R.~Espinosa, G.~F.~Giudice, E.~Morgante, A.~Riotto, L.~Senatore, A.~Strumia and N.~Tetradis,
  %``The cosmological Higgstory of the vacuum instability,''
  JHEP {\bf 1509} (2015) 174.
  %%CITATION = doi:10.1007/JHEP09(2015)174;%%

 %\cite{Hook:2014uia}
\bibitem{Hook:2014uia} 
  A.~Hook, J.~Kearney, B.~Shakya and K.~M.~Zurek,
  %``Probable or Improbable Universe? Correlating Electroweak Vacuum Instability with the Scale of Inflation,''
  JHEP {\bf 1501}, 061 (2015);
%  [arXiv:1404.5953 [hep-ph]]
  %%CITATION = ARXIV:1404.5953;%%
  %
  %\cite{Kearney:2015vba}
%\bibitem{Kearney:2015vba} 
 J.~Kearney, H.~Yoo and K.~M.~Zurek,
  %``Is a Higgs Vacuum Instability Fatal for High-Scale Inflation?,''
  Phys.\ Rev.\ D {\bf 91} (2015) 12,  123537.
  %%CITATION = doi:10.1103/PhysRevD.91.123537;%%
 

%\cite{Fairbairn:2014zia}
\bibitem{Fairbairn:2014zia} 
 M.~Fairbairn and R.~Hogan,
 %``Electroweak Vacuum Stability in light of BICEP2,''
 Phys.\ Rev.\ Lett.\ {\bf 112}, 201801 (2014).
 %%CITATION = ARXIV:1403.6786;%%
 %31 citations counted in INSPIRE as of 15 juin 2015
 
 %\cite{Shkerin:2015exa}
\bibitem{Shkerin:2015exa} 
 A.~Shkerin and S.~Sibiryakov,
 %``On stability of electroweak vacuum during inflation,''
 Phys.\ Lett.\ B {\bf 746}, 257 (2015);
 %%CITATION = ARXIV:1503.02586;%%
 %6 citations counted in INSPIRE as of 15 juin 2015
A.~Kobakhidze and A.~Spencer-Smith,
 %``Electroweak Vacuum (In)Stability in an Inflationary Universe,''
 Phys.\ Lett.\ B {\bf 722}, 130 (2013).
 %%CITATION = ARXIV:1301.2846;%%

%\cite{Kamada:2014ufa}
\bibitem{Kamada:2014ufa} 
 K.~Kamada,
 %``Inflationary cosmology and the standard model Higgs with a small Hubble induced mass,''
 Phys.\ Lett.\ B {\bf 742}, 126 (2015).
 %%CITATION = ARXIV:1409.5078;%%
 %9 citations counted in INSPIRE as of 15 Jun 2015

%\cite{Altarelli:1994rb}
\bibitem{Altarelli:1994rb} 
 G.~Altarelli and G.~Isidori,
 %``Lower limit on the Higgs mass in the standard model: An Update,''
 Phys.\ Lett.\ B {\bf 337}, 141 (1994).
 %%CITATION = PHLTA,B337,141;%%
 %338 citations counted in INSPIRE as of 31 May 2015
 
 %\cite{Linde:1983gd}
\bibitem{Linde:1983gd} 
  A.~D.~Linde,
  %``Chaotic Inflation,''
  Phys.\ Lett.\ B {\bf 129}, 177 (1983).
  %%CITATION = PHLTA,B129,177;%%

%\cite{Enqvist:2013eua}
\bibitem{Enqvist:2013eua} 
 K.~Enqvist and M.~Karciauskas,
 %``Does Planck really rule out monomial inflation?,''
 JCAP {\bf 1402}, 034 (2014).
 %%CITATION = ARXIV:1312.5944;%%
 %9 citations counted in INSPIRE as of 31 May 2015

%\cite{Coleman:1973jx}
\bibitem{Coleman:1973jx} 
 S.~R.~Coleman and E.~J.~Weinberg,
 %``Radiative Corrections as the Origin of Spontaneous Symmetry Breaking,''
 Phys.\ Rev.\ D {\bf 7}, 1888 (1973).
 %%CITATION = PHRVA,D7,1888;%%
 %3386 citations counted in INSPIRE as of 31 May 2015

%\cite{Kofman:1997yn}
\bibitem{Kofman:1997yn} 
 L.~Kofman, A.~D.~Linde and A.~A.~Starobinsky,
 %``Towards the theory of reheating after inflation,''
 Phys.\ Rev.\ D {\bf 56}, 3258 (1997);
 %%CITATION = HEP-PH/9704452;%%
 %904 citations counted in INSPIRE as of 12 juin 2015
P.~B.~Greene and L.~Kofman,
 %``On the theory of fermionic preheating,''
 Phys.\ Rev.\ D {\bf 62}, 123516 (2000).
 
 %\cite{Allahverdi:2010xz}
\bibitem{Allahverdi:2010xz} 
  R.~Allahverdi, R.~Brandenberger, F.~Y.~Cyr-Racine and A.~Mazumdar,
  %``Reheating in Inflationary Cosmology: Theory and Applications,''
  Ann.\ Rev.\ Nucl.\ Part.\ Sci.\  {\bf 60}, 27 (2010).
%  [arXiv:1001.2600 [hep-th]].
  %%CITATION = ARXIV:1001.2600;%%

%\cite{Fukugita:1986hr}
\bibitem{Fukugita:1986hr} 
 M.~Fukugita and T.~Yanagida,
 %``Baryogenesis Without Grand Unification,''
 Phys.\ Lett.\ B {\bf 174}, 45 (1986).
 %%CITATION = PHLTA,B174,45;%%
 %2380 citations counted in INSPIRE as of 02 Jun 2015

%\cite{Lyth:1998xn}
\bibitem{Lyth:1998xn} 
 D.~H.~Lyth and A.~Riotto,
 %``Particle physics models of inflation and the cosmological density perturbation,''
 Phys.\ Rept.\ {\bf 314}, 1 (1999).
 %%CITATION = HEP-PH/9807278;%%
 %1312 citations counted in INSPIRE as of 02 juin 2015

%\cite{Linde:2005ht}
\bibitem{Linde:2005ht} 
 A.~D.~Linde,
 %``Particle physics and inflationary cosmology,''
 Contemp.\ Concepts Phys.\ {\bf 5}, 1 (1990).
 %[hep-th/0503203].
 %%CITATION = HEP-TH/0503203;%%
 %423 citations counted in INSPIRE as of 04 juin 2015

%\cite{Lebedev:2012zw}
\bibitem{Lebedev:2012zw} 
 O.~Lebedev,
 %``On Stability of the Electroweak Vacuum and the Higgs Portal,''
 Eur.\ Phys.\ J.\ C {\bf 72}, 2058 (2012).
 %%CITATION = ARXIV:1203.0156;%%
 %80 citations counted in INSPIRE as of 15 juin 2015

\end{thebibliography}
\end{document}